\documentclass[paper,notoc]{JHEP3}
\usepackage{epsfig,cite}
\usepackage{amsbsy} 
\usepackage{graphics}
\def\beq{\begin{equation}}   
\def\eeq{\end{equation}}
\def\bea{\begin{eqnarray}}  
\def\eea{\end{eqnarray}}

\def\ihixs{{\tt iHixs}}
\title{Inclusive Higgs boson cross-section  for the LHC at 8 TeV}

\author{Charalampos Anastasiou\\
  Institute for Theoretical Physics, ETH Zurich,
  8093 Zurich, Switzerland\\
  E-mail: \email{babis@phys.ethz.ch}}

\author{Stephan Buehler\\
Institute for Theoretical Physics, ETH Zurich,
  8093 Zurich, Switzerland\\
  E-mail: \email{buehler@itp.phys.ethz.ch}}

\author{Franz Herzog\\
  Institute for Theoretical Physics, ETH Zurich,
  8093 Zurich, Switzerland\\
  E-mail: \email{fherzog@itp.phys.ethz.ch}}

\author{Achilleas Lazopoulos\\
  Institute for Theoretical Physics, ETH Zurich,
  8093 Zurich, Switzerland\\
  E-mail: \email{lazopoli@itp.phys.ethz.ch}}
\abstract{
We present the inclusive Higgs boson cross-section at the LHC with collision energy of 
8 TeV. Our predictions  are obtained using our publicly available program {\ihixs} 
which incorporates NNLO QCD corrections and electroweak corrections.  We review the 
convergence of the QCD perturbative expansion at this new energy and examine  
the impact of finite Higgs width effects.  We  also study the impact of different parton 
distribution functions on the cross-section. We present tables with the cross-section 
values and estimates for their uncertainty due to uncalculated  higher orders in the perturbative 
expansion and parton densities.   
}
\begin{document}
\section{Introduction}

Experiments at the Large  Hadron Collider have  made an impressive progress in the search for the Higgs  boson during 2011. 
In the Standard Model, only a small window of Higgs boson masses is in agreement 
with LHC~\cite{Collaboration:2012si,Chatrchyan:2012tx}, Tevatron~\cite{Aaltonen:2011gs} and LEP~\cite{Barate:2003sz} data.  
The  search for the Higgs boson will resume shortly in 2012.  A discovery or exclusion of  a Standard Model Higgs is eminent,  
provided of course  that the theoretical prediction is solid and that the LHC machine  and experiments  perform as anticipated.  
In 2012, proton-proton collisions at  the LHC will have a new center of mass energy  of 8 TeV. 

The purpose of this article  is to provide numerical results for the inclusive gluon fusion Higgs boson cross-section at 8 TeV. 
We obtain state of the art  predictions for the inclusive gluon fusion
cross section and its uncertainties with our publicly available 
computer program {\ihixs}~\cite{ihixs}. 
{\ihixs\,} evaluates the contribution to the cross-section in NNLO QCD
and includes important electroweak effects.  
A detailed description of the theoretical 
contributions~\cite{Anastasiou:2002yz,Anastasiou:2006hc,Anastasiou:2009kn,Furlan:2011uq, Buehler:2011ev,Aglietti:2004nj,Anastasiou:2008tj,Actis:2008ug,vanHameren:2010cp,Ellis:2007qk,Hahn:2004fe,Whalley:2005nh,Harlander:2002wh,Djouadi:1997yw,restwork} which are incorporated and accounted for in {\ihixs} can be found in the corresponding 
publication ~\cite{ihixs}.

In Section~\ref{sec:convergence} we  study the convergence of the perturbative QCD corrections. 
In Section~\ref{sec:pdfs} we  study the sensitivity of the
cross-section on parton densities.  
In Section~\ref{sec:width} we study the effect of the Higgs width.  
In Section~\ref{sec:tables} we  present our numerical 
values  for the cross-section and its uncertainties.  

\section{Perturbative convergence and scale uncertainties}
\label{sec:convergence}

\begin{figure}[h]
\begin{minipage}[b]{0.5\linewidth}
\includegraphics[width=\linewidth]{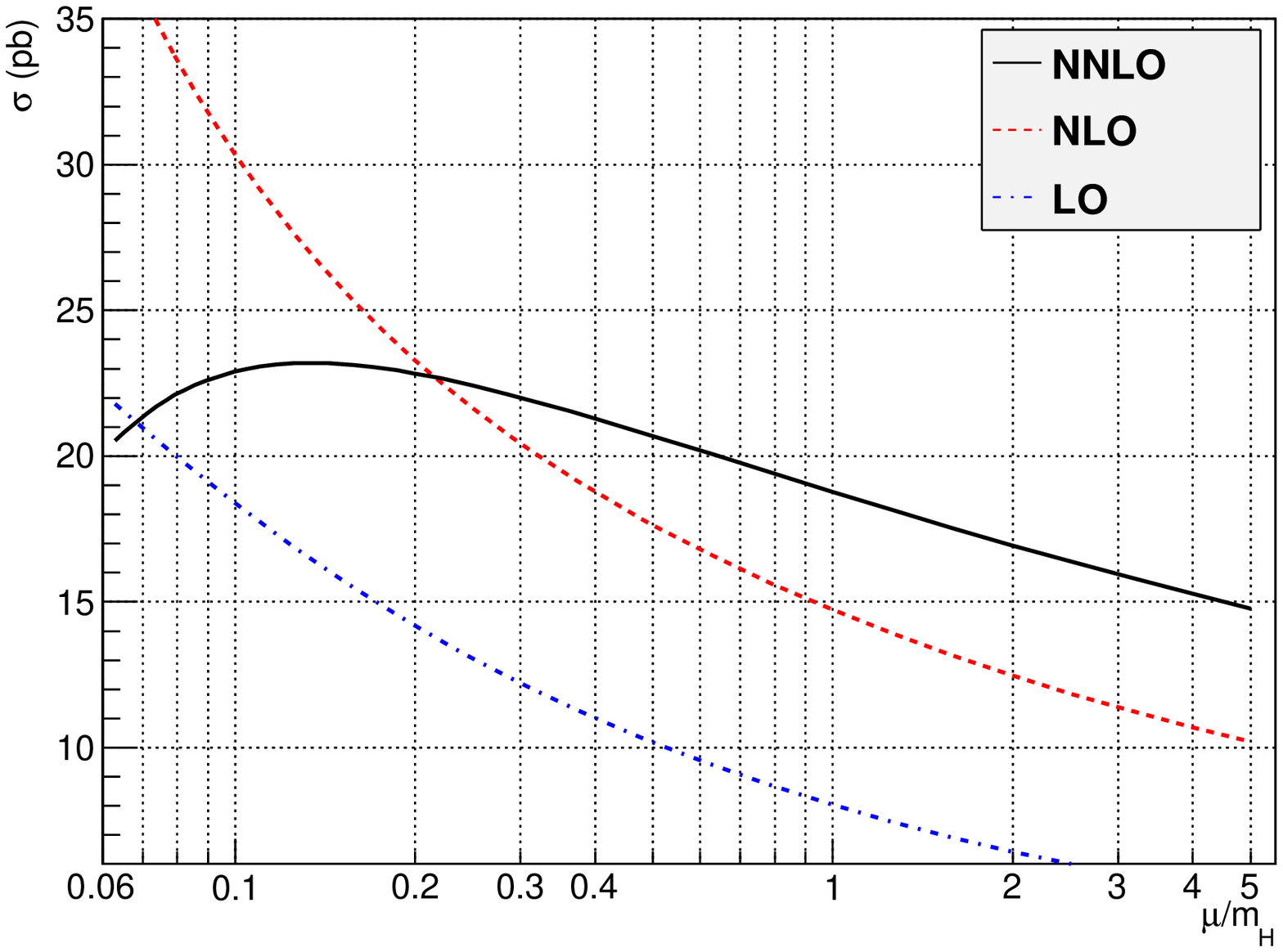}
\end{minipage}
\begin{minipage}[b]{0.5\linewidth}
\includegraphics[width=\linewidth]{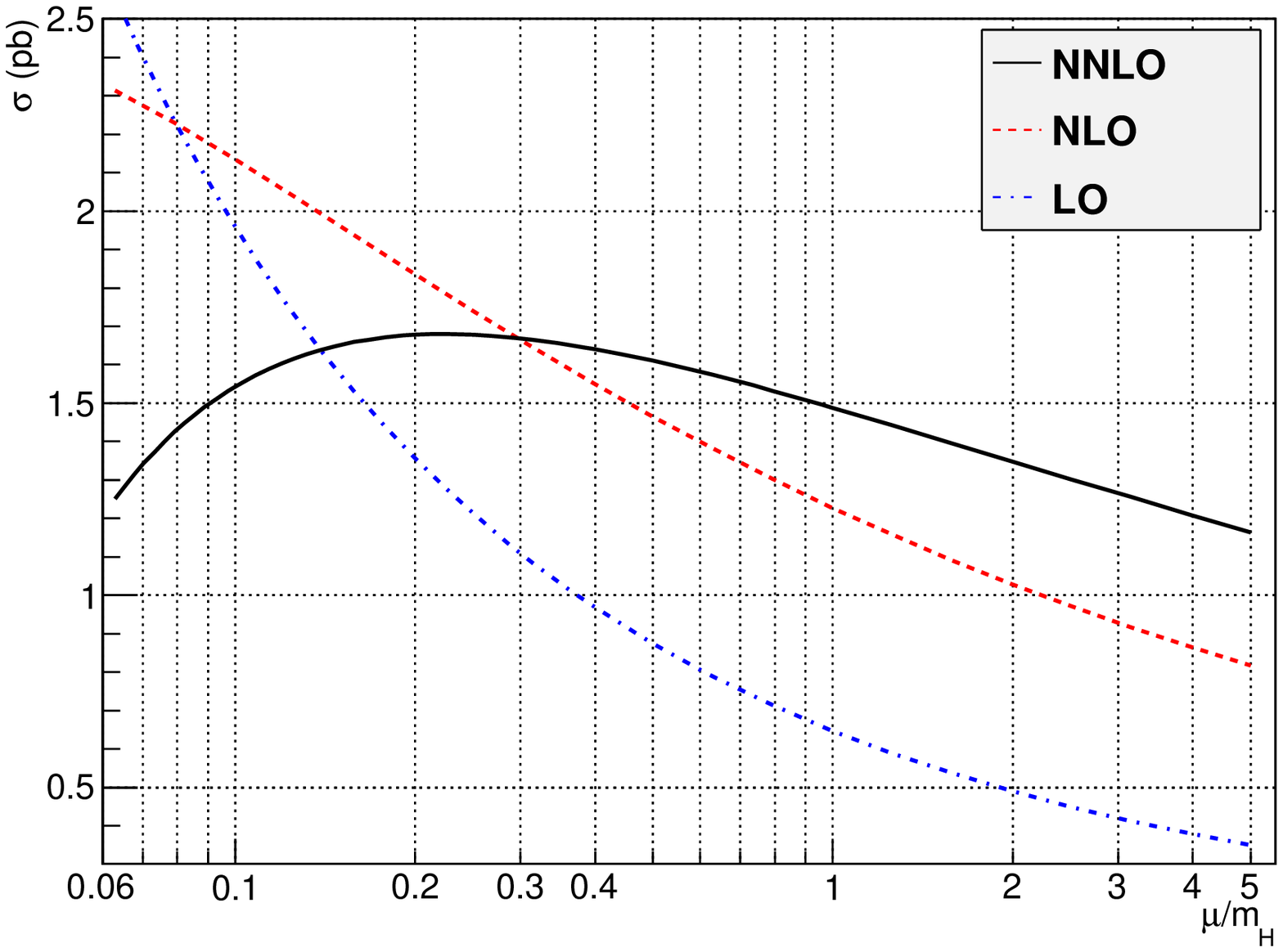}
\end{minipage}
\label{fig:scalevariation}
\caption{Scale dependence of the gluon fusion cross-section at LO, NLO, and NNLO for $m_H = 125 {\rm GeV}$ (left panel) and  $m_H = 450 {\rm GeV}$ (right panel). The perturbative series converges faster  for scale choices smaller than  Higgs boson mass. }
\end{figure}
The  perturbative convergence of the Higgs boson cross-section has been studied thoroughly during the last decade.  
We  find  a  similar convergence pattern at the new LHC energy of $8\,
{\rm TeV}$ as for $7\, {\rm TeV}$ and  
$14\,{\rm TeV}$.  For illustration, we present in
Figure~\ref{fig:scalevariation} the behavior of the cross-section at
$8\,{\rm TeV}$  by varying the renormalization 
and factorization scales $\mu=\mu_R=\mu_F$ simultaneously\footnote{We use the central MSTW08 PDF set for all results in this section.}.  
As  it has already been realized in the literature, smaller scales than the Higgs boson mass 
lead to a faster convergence of the perturbative
expansion~\cite{ihixs,Anastasiou:2002yz}.  
\begin{figure}[h]
\includegraphics[width=\linewidth]{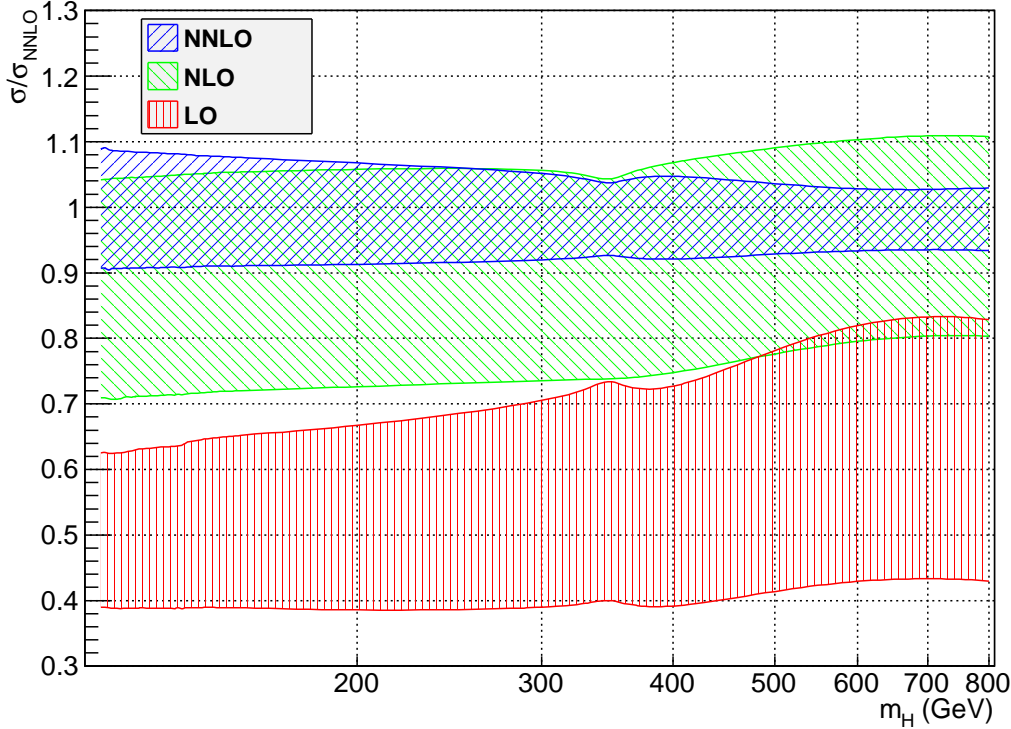}
\label{fig:scaleband}
\caption{Scale variation of the gluon fusion cross-section at LO, NLO, and NNLO. The LO, NLO and NNLO cross-sections at scale $\mu\in \left[ \frac{m_H}{4}, m_H\right]$ are normalized to the NNLO cross-section at the central scale $\mu = \frac{m_H}{2}$.} 
\end{figure}

We estimate the theoretical uncertainty from uncalculated higher order
corrections  by varying the renormalization and factorization 
scale in the interval $\mu\in \left[ \frac{m_H}{4}, m_H\right]$. 
In Fig~\ref{fig:scaleband} we  present the cross-section at LO, NLO and NNLO 
in this interval, normalized to the NNLO cross-section at the central
scale $\mu = \frac{m_H}{2}$.  The NNLO and NLO bands overlap largely,
Corrections beyond NNLO would need to be atypical in order for our
uncertainty estimate to be inaccurate.    The cross-section is known
to be stable under threshold and other corrections which can be
resummed beyond NNLO~\cite{deFlorian:2009hc,Ahrens:2010rs,Moch:2005ky}. 

\section{Parton density function comparison}
\label{sec:pdfs}

\begin{figure}[h]
\includegraphics[width=\linewidth]{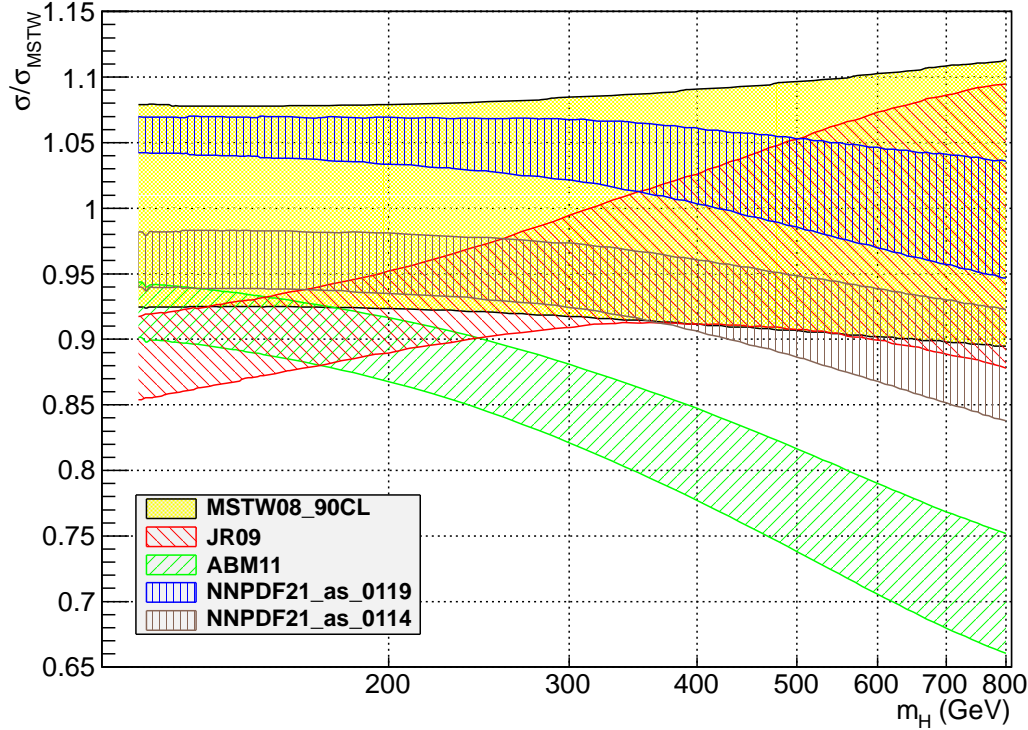}
\label{fig:pdfband}
\caption{The uncertainty of the Higgs cross-section due to the parton distribution functions.} 
\end{figure}

The Higgs boson cross-section requires parton distribution functions (pdf)
as input.  In  {\ihixs\,}  we employ all  pdf sets that allow for
NNLO evolution (MSTW08~\cite{Martin:2009iq},
JR09\cite{JimenezDelgado:2009tv},NNPDF\cite{Ball:2011uy},
ABM11\cite{Alekhin:2012ig}) and are publicly available through the LHAPDF~\cite{Whalley:2005nh,lhapdf} library.  

Our knowledge of the parton densities has been increasing steadily over the recent years.
The gluon density is being extracted  with an accuracy of a few percent over a large range  of
Bjorken-$x$ values.  However,  a direct comparison exposes 
differences beyond  the quoted uncertainties. These discrepancies
affect the gluon density and the fitted value of $\alpha_s$ which are 
very important input for the calculation of the Higgs boson
cross-section. 
  
In Figure~\ref{fig:pdfband} we  present uncertainty bands for  each
pdf provider, normalizing all cross-sections to the central value of 
the NNLO MSTW08 set.  The MSTW uncertainty band is at $90\%$
confidence level~\footnote{The $68\%$ confidence level uncertainty 
 reduces the width of the band to about half, as it is anticipated by 
a naive error propagation of the gluon density uncertainty to the
gluon partonic luminosity}.  
The uncertainties of the other pdf sets are at $68\%$ confidence
level.  For the NNPDF set we present the Higgs
cross-section for the nominal value of the strong coupling $\alpha_s=0.119$
and for the lowest value $\alpha_s = 0.114$ for which a  set is provided. This value is close to that extracted by the ABM11 group and shows that the discrepancy between the two sets cannot be attributed entirely to the value of $a_s$. 

The differences of  pdf sets beyond their quoted  uncertainties has been a
tantalizing issue. The discrepancies between the MSTW and NNPDF pdf sets (and in the high mass region, also the JR set) can be partially accounted for by enlarging the
uncertainty of the MSTW08 set to the one obtained at $90\% $ 
confidence level. We will therefore use MSTW08 at $90\%$CL as our main benchmark set. 

The discrepancy between the ABM11 and the other sets is  large and
becomes rather dramatic for high values of the Higgs boson mass.   
Notably, the ABM11 set finds a value of $\alpha_s=0.1134(11)$ which is
quite small in comparison to the world average ($\alpha_s=0.1184(7)$)~\cite{Bethke:2009jm}.  
It is alarming to us that  the extractions of $\alpha_s$ based on 
quantities which are known precisely from theory, such as the total 
$Z$ boson hadronic width~\cite{Baikov:2012er}, are in tension with the
small  value of $\alpha_s$ of the ABM11 and JR sets.  
This argument aside, the ABM11 set is not proven to be inconsistent
with LHC data. We believe that we cannot disregard it for the crucial studies in
the search of the Higgs boson. 

One could attempt to enlarge further the pdf uncertainty that we have assigned to our benchmark
MSTW08 set. Such a remedy is inadequate for the 
calculation of the likelihood of a Higgs boson mass hypothesis in LHC
data. Uncertainties 
assigned to parton densities are typically treated as nuisance
parameters in such a calculation. This can be largely justified by
the fact that a significant component of the pdf uncertainty is
due to statistical fluctuations of the measurements from which they
are extracted.  
However, the discrepancy between ABM11 and MSTW08 is of a systematic
nature. Given the large difference in central values, enlarging the
statistical uncertainty of the MSTW08 set is equivalent to arbitrarily assigning a smaller
weight to the ABM11 cross-section values. 

Until a clearer theoretical understanding of the pdf discrepancies is
available, we find it prudent to adopt the ABM11 set as our second 
benchmark set.  In this publication, we  provide tables of
cross-section values for both the MSTW08 set and the ABM11 set.  
For the MSTW08 we quote the pdf uncertainty  at the $90\%$ CL. We use the ABM11 set (with $68\%$ CL uncertainty) as a means to obtain a lower estimate of the 
cross-section.

\section{Higgs boson width and signal-background interference}
\label{sec:width}

\begin{figure}[h]
\includegraphics[width=\linewidth]{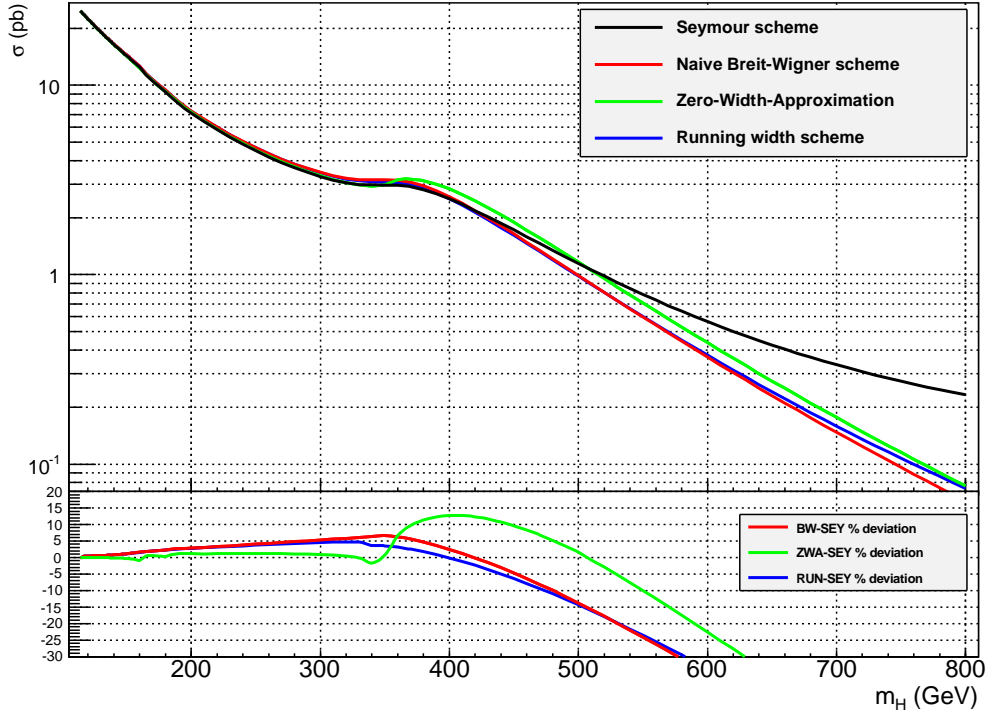}
\label{fig:width}
\caption{Inclusive Higgs cross-section at NNLO  with various treatments of the Higgs propagator. The percentage difference with respect to the cross-section in the Seymour scheme is shown in the lower panel.}
\end{figure}

{\ihixs} computes  the partonic cross-section for resonant  production and decay of a Higgs boson
in the $s$-channel approximation, 
by performing  an integration over the invariant mass of the Higgs decay products:    
\begin{equation}
\label{eq:BWG}
\hat{\sigma}_{ij \to \{H_{\rm final}\}+X}(\hat{s}, \mu_f)    = 
\int_{Q_{a}^2}^{Q_{b}^2} 
dQ^2 \frac{Q \Gamma_H(Q)}{ \pi } 
\frac{
\hat{\sigma}_{ij \to H}(\hat{s}, Q^2, \mu_f) 
{\rm Br}_{H \to \{H_{\rm final} \}}(Q)}{{\Delta}\left(Q, m_H\right)}.
\end{equation} 
The total cross-section corresponds to summing over all Higgs boson final states, 
\begin{equation}
\sum_{{\rm final}} {\rm Br}_{H \to \{H_{\rm final} \}}(Q) =1. 
\end{equation} 
The function $\Delta$ arises from s-channel Higgs propagator diagrams and it diverges at leading order  in 
perturbation theory for $Q \to m_H$.  A  resummation of dominant Feynman diagrams in this limit 
at all orders in perturbation theory is  necessary in order to obtain a sensible finite result. 

{\ihixs } permits an easy implementation of various schemes for the
propagator function $\Delta$. An optimal treatment of finite width
effects is the subject of ongoing research. As a diagnostic of
potentially sizable contributions we have employed the following
schemes: 
\begin{itemize}
\item  \textbf{Zero width approximation} (ZWA): 
\begin{equation}
\frac{1}{\Delta\left(Q, m_H\right) } \to 
\frac{\pi}{m_H \Gamma_H(m_H)} \delta\left(Q^2 -m_H^2\right)
\end{equation}
This should be an adequate treatment for a light Higgs boson where 
the width of the Higgs boson is very small in comparison to its mass.  
\item  \textbf{Naive Breit-Wigner} (BW):
\begin{equation}
\frac{1}{\Delta\left(Q, m_H\right) } = 
\frac{1}
{ (Q^2 -m_H^2)^2 + m_H^2 \Gamma_H^2(m_H)}
\end{equation}
This corresponds to resumming leading width contributions for $Q \sim
m_H$. 
\item \textbf{Breit-Wigner with running width}:  
\begin{equation}
\frac{1}{\Delta\left(Q, m_H\right) } = 
\frac{1}
{ (Q^2 -m_H^2)^2 + Q^2 \Gamma_H^2(Q)}
\end{equation}
We evaluate the width of the Higgs boson using {\tt
  HDECAY}~\cite{Djouadi:1997yw}.  Our numerical results of
Section~\ref{sec:tables} are obtained with this scheme. 
\item  \textbf{Seymour scheme} ~\cite{Seymour:1995qg}  
\begin{equation}
\frac{1}{\Delta(Q,m_H)} = \frac{ m_H^4 / Q^4}{(Q^2-m_H^2)^2
+\Gamma^2_H(m_H^2)\frac{Q^4}{m_H^2}}
\end{equation}
This scheme is a prescription to account for signal-background  interference effects at the high 
energy limit and finite width resummation simultaneously.  
\end{itemize}

In Figure~\ref{fig:width}, we present the Higgs boson cross-section
using the above treatments of Higgs width effects as  a function of the Higgs
boson mass. We notice that finite width  
and  signal-background interference effects (as accounted for in the Seymour
scheme~\cite{Seymour:1995qg})  become important for Higgs boson masses above $400\,{\rm GeV}$.  
For such high values of the Higgs boson mass, where the width grows,  
the precise form of the Higgs propagator $\Delta(Q, m_H)$ is important and requires a consistent resummation of resonant
diagrams. For a recent study we refer  the reader to
Ref.~\cite{Goria:2011wa}.   We emphasize that signal-background
interference effects must always be accounted for by dedicated
calculations such as in Refs~\cite{Dixon:2003yb,Dixon:2008xc,Duhrssen:2005bz,Binoth:2006mf,Campbell:2011cu}.

\section{Numerical estimates of the gluon fusion cross-section}

In this Section, we present numerical results for the Higgs boson
total cross-section at the LHC with 8 \rm {TeV} energy  and values of
the Higgs boson mass in the range $m_H \in [114, 400]\,  {\rm GeV}$. 
We treat finite width effects using the running width scheme of Section~\ref{sec:width}.
We defer a detailed study of the case of a heavier Higgs boson for a
future publication.  As explained in Section~\ref{sec:pdfs} we present, in tables~\ref{xsection1} and~\ref{xsection2}, 
results for two pdf benchmark sets: a representative set MSTW08 (with pdf
uncertainties quoted at the $90\%$ confidence level) and  the ABM11 set
which typically yields the smallest Higgs boson cross-sections (with
pdf uncertainties quoted at the $68\%$ confidence level). The
predictions of these two sets are discrepant and we do not attempt to
combine them in an ``average''  of any type. 

We note that all of the results in this section can be reproduced with
the publicly available program {\ihixs}.   The runcard options that
we have used to obtain the central values of the cross-section
predictions are: 
\begin{center}
\fbox{
\begin{minipage}[b]{0.69\textwidth}

{\tt pdf\_provider : \{MSTW90,MSTW90\_M,MSTW90\_P,ABM\} \\
effective\_theory\_flag = 0 \\
no\_error\_flag = 0\\
collider = LHC\\
Etot = 8000\\
higgs\_width\_scheme = 2\\
mhiggs  : [114,400]\\
muf/mhiggs  : \{0.5,0.25,1.0\}\\
mur/mhiggs  : \{0.5,0.25,1.0\}\\
DecayMode  = total\\
ProductionMode  = gg\\
K\_ewk  = 1.0
}
\end{minipage}
\begin{minipage}[b]{0.29\textwidth}

{\tt

K\_ewk\_real  = 1.0\\
K\_ewk\_real\_b  = 1.0\\
m\_top  = 172.5\\
Gamma\_top  = 0.0\\
Y\_top  = 1.0\\
m\_bot  = 3.63\\
Gamma\_bot  = 0.0\\
Y\_bot  = 1.0\\
m\_Z  = 91.1876\\
Gamma\_Z  = 2.4952\\
m\_W  = 80.403\\
Gamma\_W  = 2.141
}
\end{minipage}
}

\end{center}

\section{Summary}
\label{sec:conclusions}
In this article, we have presented predictions for the inclusive gluon
fusion cross-section at the LHC with 8 {\rm TeV} center of mass
energy using our publicly available program \ihixs~\cite{ihixs}.  
We have reviewed the perturbative convergence of the
cross-section and derived uncertainty estimates due to
uncalculated higher order effects by means of scale variations. 

We have also reviewed the sensitivity of the cross-section on all 
available NNLO sets of parton distribution functions which are
available through LHAPDF~\cite{lhapdf}.  
We have adopted the MSTW08
pdf set with uncertainties at the $90\%$ confidence level 
as our representative benchmark.  We have observed that the  
ABM11 pdf set typically yields the smallest cross-section values.  
We present the cross-section predictions with ABM11 which can be used for a more conservative estimate of the Higgs boson production rate.  

We  have studied the impact of finite width effects for large values
of the Higgs boson mass. We  limit our predictions  to mass values
where these effects can be neglected.  The case of very heavy and very
wide Higgs boson will be analyzed in a future publication. 

We hope that our results will be a useful input for the
experimental searches of the Higgs boson at the LHC 
in 2012.

\acknowledgments
Research supported by the Swiss National Foundation under contract SNF 200020-126632.

\newpage
\label{sec:tables}

\begin{table}[h]
\begin{center}
 \begin{tabular}{| c ||  c | c | c || c|c|c|}
\hline
$m_H$(GeV)	&MSTW08 $\sigma(pb)$  & $\% \delta_{PDF}$ & $\%  \delta_{\mu_F}$ 
	   	&   ABM11 $\sigma(pb)$  &  $\% \delta_{PDF}$ &$\% \delta_{\mu_F}$\\
\hline\hline
114 & 24.69& ${}^{+7.92}_{-7.54}$ & ${}^{+8.83}_{-9.32}$  & 22.78& ${}^{+2.28}_{-2.28}$ & ${}^{+8.0}_{-8.85}$ \\ \hline
115 & 24.27& ${}^{+7.91}_{-7.54}$ & ${}^{+9.07}_{-9.31}$  & 22.38& ${}^{+2.29}_{-2.29}$ & ${}^{+7.98}_{-8.84}$ \\ \hline
116 & 23.94& ${}^{+7.9}_{-7.61}$ & ${}^{+8.75}_{-9.59}$  & 22.0& ${}^{+2.29}_{-2.29}$ & ${}^{+8.0}_{-8.83}$ \\ \hline
117 & 23.55& ${}^{+7.93}_{-7.54}$ & ${}^{+8.64}_{-9.33}$  & 21.68& ${}^{+2.29}_{-2.29}$ & ${}^{+7.92}_{-9.05}$ \\ \hline
118 & 23.17& ${}^{+7.92}_{-7.54}$ & ${}^{+8.6}_{-9.38}$  & 21.33& ${}^{+2.3}_{-2.3}$ & ${}^{+7.84}_{-8.84}$ \\ \hline
119 & 22.79& ${}^{+7.92}_{-7.53}$ & ${}^{+8.55}_{-9.35}$  & 20.98& ${}^{+2.3}_{-2.3}$ & ${}^{+7.79}_{-8.87}$ \\ \hline
120 & 22.42& ${}^{+7.91}_{-7.53}$ & ${}^{+8.53}_{-9.3}$  & 20.63& ${}^{+2.3}_{-2.3}$ & ${}^{+7.77}_{-8.85}$ \\ \hline
121 & 22.06& ${}^{+7.91}_{-7.53}$ & ${}^{+8.51}_{-9.34}$  & 20.29& ${}^{+2.3}_{-2.3}$ & ${}^{+7.75}_{-8.82}$ \\ \hline
122 & 21.7& ${}^{+7.91}_{-7.53}$ & ${}^{+8.47}_{-9.28}$  & 19.96& ${}^{+2.31}_{-2.31}$ & ${}^{+7.74}_{-8.82}$ \\ \hline
123 & 21.36& ${}^{+7.8}_{-7.53}$ & ${}^{+8.42}_{-9.28}$  & 19.64& ${}^{+2.31}_{-2.31}$ & ${}^{+7.72}_{-8.86}$ \\ \hline
124 & 21.02& ${}^{+7.81}_{-7.52}$ & ${}^{+8.41}_{-9.25}$  & 19.32& ${}^{+2.31}_{-2.31}$ & ${}^{+7.68}_{-8.81}$ \\ \hline
125 & 20.69& ${}^{+7.79}_{-7.53}$ & ${}^{+8.37}_{-9.26}$  & 19.01& ${}^{+2.32}_{-2.32}$ & ${}^{+7.65}_{-8.82}$ \\ \hline
126 & 20.37& ${}^{+7.8}_{-7.53}$ & ${}^{+8.35}_{-9.24}$  & 18.71& ${}^{+2.32}_{-2.32}$ & ${}^{+7.64}_{-8.8}$ \\ \hline
127 & 20.05& ${}^{+7.8}_{-7.52}$ & ${}^{+8.34}_{-9.21}$  & 18.41& ${}^{+2.32}_{-2.32}$ & ${}^{+7.6}_{-8.84}$ \\ \hline
128 & 19.74& ${}^{+7.79}_{-7.52}$ & ${}^{+8.3}_{-9.2}$  & 18.13& ${}^{+2.33}_{-2.33}$ & ${}^{+7.58}_{-8.79}$ \\ \hline
129 & 19.44& ${}^{+7.8}_{-7.52}$ & ${}^{+8.28}_{-9.26}$  & 17.84& ${}^{+2.33}_{-2.33}$ & ${}^{+7.56}_{-8.79}$ \\ \hline
130 & 19.14& ${}^{+7.79}_{-7.51}$ & ${}^{+8.24}_{-9.19}$  & 17.57& ${}^{+2.33}_{-2.33}$ & ${}^{+7.54}_{-8.84}$ \\ \hline
131 & 18.86& ${}^{+7.8}_{-7.51}$ & ${}^{+8.22}_{-9.17}$  & 17.3& ${}^{+2.34}_{-2.34}$ & ${}^{+7.51}_{-8.79}$ \\ \hline
132 & 18.57& ${}^{+7.79}_{-7.51}$ & ${}^{+8.19}_{-9.16}$  & 17.03& ${}^{+2.34}_{-2.34}$ & ${}^{+7.47}_{-8.77}$ \\ \hline
133 & 18.3& ${}^{+7.8}_{-7.5}$ & ${}^{+8.17}_{-9.15}$  & 16.77& ${}^{+2.35}_{-2.35}$ & ${}^{+7.46}_{-8.75}$ \\ \hline
134 & 18.03& ${}^{+7.79}_{-7.51}$ & ${}^{+8.14}_{-9.15}$  & 16.52& ${}^{+2.35}_{-2.35}$ & ${}^{+7.41}_{-8.74}$ \\ \hline
135 & 17.76& ${}^{+7.8}_{-7.51}$ & ${}^{+8.12}_{-9.19}$  & 16.27& ${}^{+2.35}_{-2.35}$ & ${}^{+7.39}_{-8.73}$ \\ \hline
136 & 17.5& ${}^{+7.8}_{-7.5}$ & ${}^{+8.05}_{-9.17}$  & 16.03& ${}^{+2.36}_{-2.36}$ & ${}^{+7.37}_{-8.73}$ \\ \hline
137 & 17.25& ${}^{+7.78}_{-7.53}$ & ${}^{+8.05}_{-9.17}$  & 15.79& ${}^{+2.36}_{-2.36}$ & ${}^{+7.36}_{-8.75}$ \\ \hline
138 & 17.01& ${}^{+7.79}_{-7.51}$ & ${}^{+8.01}_{-9.13}$  & 15.56& ${}^{+2.37}_{-2.37}$ & ${}^{+7.31}_{-8.73}$ \\ \hline
139 & 16.77& ${}^{+7.8}_{-7.51}$ & ${}^{+7.97}_{-9.08}$  & 15.34& ${}^{+2.37}_{-2.37}$ & ${}^{+7.26}_{-8.7}$ \\ \hline
140 & 16.53& ${}^{+7.79}_{-7.5}$ & ${}^{+7.9}_{-9.06}$  & 15.12& ${}^{+2.37}_{-2.37}$ & ${}^{+7.24}_{-8.69}$ \\ \hline
141 & 16.3& ${}^{+7.79}_{-7.5}$ & ${}^{+7.88}_{-9.03}$  & 14.9& ${}^{+2.38}_{-2.38}$ & ${}^{+7.23}_{-8.67}$ \\ \hline
142 & 16.07& ${}^{+7.79}_{-7.5}$ & ${}^{+7.87}_{-9.01}$  & 14.69& ${}^{+2.38}_{-2.38}$ & ${}^{+7.2}_{-8.62}$ \\ \hline
143 & 15.85& ${}^{+7.78}_{-7.51}$ & ${}^{+7.85}_{-9.0}$  & 14.48& ${}^{+2.39}_{-2.39}$ & ${}^{+7.19}_{-8.62}$ \\ \hline
144 & 15.64& ${}^{+7.78}_{-7.5}$ & ${}^{+7.82}_{-8.99}$  & 14.28& ${}^{+2.39}_{-2.39}$ & ${}^{+7.18}_{-8.62}$ \\ \hline
145 & 15.43& ${}^{+7.78}_{-7.51}$ & ${}^{+7.79}_{-8.99}$  & 14.08& ${}^{+2.4}_{-2.4}$ & ${}^{+7.16}_{-8.6}$ \\ \hline
146 & 15.22& ${}^{+7.79}_{-7.51}$ & ${}^{+7.78}_{-8.97}$  & 13.88& ${}^{+2.4}_{-2.4}$ & ${}^{+7.18}_{-8.57}$ \\ \hline
147 & 15.02& ${}^{+7.79}_{-7.5}$ & ${}^{+7.74}_{-8.97}$  & 13.69& ${}^{+2.41}_{-2.41}$ & ${}^{+7.14}_{-8.59}$ \\ \hline
148 & 14.81& ${}^{+7.8}_{-7.51}$ & ${}^{+7.74}_{-8.97}$  & 13.5& ${}^{+2.41}_{-2.41}$ & ${}^{+7.12}_{-8.58}$ \\ \hline
149 & 14.62& ${}^{+7.8}_{-7.5}$ & ${}^{+7.74}_{-8.93}$  & 13.32& ${}^{+2.42}_{-2.42}$ & ${}^{+7.1}_{-8.57}$ \\ \hline
150 & 14.43& ${}^{+7.78}_{-7.51}$ & ${}^{+7.7}_{-8.93}$  & 13.14& ${}^{+2.42}_{-2.42}$ & ${}^{+7.08}_{-8.55}$ \\ \hline
\end{tabular}
\end{center}
\caption{Inclusive Higgs production cross-section through gluon fusion (in $pb$) at $\sqrt{s}=8$ TeV, with pdf and scale uncertainties for the MSTW08 and ABM11 pdf sets. The pdf uncertainty for MSTW08 is calculated using the $90\%$CL grids, for reasons explained in section~\ref{sec:pdfs}, while the ABM11 uncertainty corresponds to $68\%$CL.  }
\label{xsection1}
\end{table}

\begin{table}[h]
\begin{center}
 \begin{tabular}{| c ||  c | c | c || c|c|c|}
\hline
$m_H$(GeV)		&MSTW08 $\sigma(pb)$  & $\% \delta_{PDF}$ & $\%  \delta_{\mu_F}$ 
	   	&   ABM11 $\sigma(pb)$  &  $\% \delta_{PDF}$ &$\% \delta_{\mu_F}$\\
\hline\hline
151 & 14.24& ${}^{+7.79}_{-7.52}$ & ${}^{+7.67}_{-8.95}$  & 12.97& ${}^{+2.43}_{-2.43}$ & ${}^{+7.04}_{-8.61}$ \\ \hline
152 & 14.06& ${}^{+7.78}_{-7.52}$ & ${}^{+7.62}_{-9.01}$  & 12.8& ${}^{+2.43}_{-2.43}$ & ${}^{+7.02}_{-8.62}$ \\ \hline
153 & 13.88& ${}^{+7.8}_{-7.52}$ & ${}^{+7.61}_{-8.99}$  & 12.63& ${}^{+2.44}_{-2.44}$ & ${}^{+7.0}_{-8.62}$ \\ \hline
154 & 13.71& ${}^{+7.8}_{-7.52}$ & ${}^{+7.6}_{-8.99}$  & 12.46& ${}^{+2.44}_{-2.44}$ & ${}^{+6.99}_{-8.6}$ \\ \hline
155 & 13.53& ${}^{+7.8}_{-7.51}$ & ${}^{+7.58}_{-8.96}$  & 12.3& ${}^{+2.45}_{-2.45}$ & ${}^{+6.98}_{-8.59}$ \\ \hline
160 & 12.66& ${}^{+7.81}_{-7.51}$ & ${}^{+7.5}_{-8.9}$  & 11.48& ${}^{+2.48}_{-2.48}$ & ${}^{+6.91}_{-8.55}$ \\ \hline
165 & 11.56& ${}^{+7.81}_{-7.54}$ & ${}^{+7.39}_{-8.87}$  & 10.47& ${}^{+2.5}_{-2.5}$ & ${}^{+6.81}_{-8.53}$ \\ \hline
170 & 10.72& ${}^{+7.83}_{-7.55}$ & ${}^{+7.26}_{-8.86}$  & 9.68& ${}^{+2.53}_{-2.53}$ & ${}^{+6.7}_{-8.52}$ \\ \hline
175 & 10.03& ${}^{+7.84}_{-7.56}$ & ${}^{+7.17}_{-8.83}$  & 9.05& ${}^{+2.57}_{-2.57}$ & ${}^{+6.64}_{-8.49}$ \\ \hline
180 & 9.42& ${}^{+7.85}_{-7.58}$ & ${}^{+7.08}_{-8.81}$  & 8.47& ${}^{+2.6}_{-2.6}$ & ${}^{+6.57}_{-8.47}$ \\ \hline
185 & 8.77& ${}^{+7.87}_{-7.59}$ & ${}^{+7.0}_{-8.78}$  & 7.88& ${}^{+2.63}_{-2.63}$ & ${}^{+6.49}_{-8.46}$ \\ \hline
190 & 8.21& ${}^{+7.87}_{-7.62}$ & ${}^{+6.91}_{-8.76}$  & 7.36& ${}^{+2.66}_{-2.66}$ & ${}^{+6.42}_{-8.45}$ \\ \hline
195 & 7.74& ${}^{+7.9}_{-7.64}$ & ${}^{+6.83}_{-8.73}$  & 6.92& ${}^{+2.7}_{-2.7}$ & ${}^{+6.33}_{-8.43}$ \\ \hline
200 & 7.32& ${}^{+7.91}_{-7.66}$ & ${}^{+6.76}_{-8.7}$  & 6.53& ${}^{+2.73}_{-2.73}$ & ${}^{+6.26}_{-8.41}$ \\ \hline
210 & 6.6& ${}^{+7.95}_{-7.71}$ & ${}^{+6.58}_{-8.65}$  & 5.86& ${}^{+2.8}_{-2.8}$ & ${}^{+6.14}_{-8.36}$ \\ \hline
220 & 6.0& ${}^{+7.98}_{-7.76}$ & ${}^{+6.42}_{-8.6}$  & 5.3& ${}^{+2.87}_{-2.87}$ & ${}^{+6.01}_{-8.32}$ \\ \hline
230 & 5.49& ${}^{+8.02}_{-7.78}$ & ${}^{+6.28}_{-8.54}$  & 4.83& ${}^{+2.95}_{-2.95}$ & ${}^{+5.87}_{-8.29}$ \\ \hline
240 & 5.05& ${}^{+8.07}_{-7.89}$ & ${}^{+6.14}_{-8.47}$  & 4.42& ${}^{+3.03}_{-3.03}$ & ${}^{+5.76}_{-8.24}$ \\ \hline
250 & 4.67& ${}^{+8.11}_{-7.9}$ & ${}^{+6.0}_{-8.41}$  & 4.07& ${}^{+3.1}_{-3.1}$ & ${}^{+5.65}_{-8.17}$ \\ \hline
260 & 4.34& ${}^{+8.15}_{-8.01}$ & ${}^{+5.84}_{-8.34}$  & 3.77& ${}^{+3.19}_{-3.19}$ & ${}^{+5.52}_{-8.11}$ \\ \hline
270 & 4.06& ${}^{+8.2}_{-8.04}$ & ${}^{+5.68}_{-8.28}$  & 3.51& ${}^{+3.27}_{-3.27}$ & ${}^{+5.37}_{-8.06}$ \\ \hline
280 & 3.82& ${}^{+8.25}_{-8.15}$ & ${}^{+5.52}_{-8.21}$  & 3.29& ${}^{+3.35}_{-3.35}$ & ${}^{+5.23}_{-8.0}$ \\ \hline
290 & 3.62& ${}^{+8.41}_{-8.16}$ & ${}^{+5.34}_{-8.13}$  & 3.09& ${}^{+3.43}_{-3.43}$ & ${}^{+5.07}_{-7.94}$ \\ \hline
300 & 3.45& ${}^{+8.46}_{-8.28}$ & ${}^{+5.14}_{-8.05}$  & 2.93& ${}^{+3.51}_{-3.51}$ & ${}^{+4.89}_{-7.85}$ \\ \hline
310 & 3.3& ${}^{+8.51}_{-8.29}$ & ${}^{+4.92}_{-7.93}$  & 2.8& ${}^{+3.6}_{-3.6}$ & ${}^{+4.69}_{-7.76}$ \\ \hline
320 & 3.2& ${}^{+8.55}_{-8.41}$ & ${}^{+4.65}_{-7.81}$  & 2.69& ${}^{+3.68}_{-3.68}$ & ${}^{+4.45}_{-7.64}$ \\ \hline
330 & 3.13& ${}^{+8.6}_{-8.42}$ & ${}^{+4.31}_{-7.65}$  & 2.62& ${}^{+3.76}_{-3.76}$ & ${}^{+4.12}_{-7.49}$ \\ \hline
340 & 3.09& ${}^{+8.65}_{-8.53}$ & ${}^{+3.87}_{-7.43}$  & 2.58& ${}^{+3.84}_{-3.84}$ & ${}^{+3.69}_{-7.28}$ \\ \hline
350 & 3.07& ${}^{+8.71}_{-8.55}$ & ${}^{+3.74}_{-7.37}$  & 2.55& ${}^{+3.92}_{-3.92}$ & ${}^{+3.59}_{-7.22}$ \\ \hline
360 & 3.06& ${}^{+8.75}_{-8.57}$ & ${}^{+4.13}_{-7.63}$  & 2.53& ${}^{+4.0}_{-4.0}$ & ${}^{+4.03}_{-7.46}$ \\ \hline
370 & 2.99& ${}^{+8.81}_{-8.69}$ & ${}^{+4.46}_{-7.79}$  & 2.46& ${}^{+4.08}_{-4.08}$ & ${}^{+4.37}_{-7.64}$ \\ \hline
380 & 2.86& ${}^{+8.86}_{-8.71}$ & ${}^{+4.65}_{-7.89}$  & 2.34& ${}^{+4.16}_{-4.16}$ & ${}^{+4.58}_{-7.75}$ \\ \hline
390 & 2.69& ${}^{+9.03}_{-8.82}$ & ${}^{+4.73}_{-7.92}$  & 2.2& ${}^{+4.24}_{-4.24}$ & ${}^{+4.67}_{-7.77}$ \\ \hline
400 & 2.51& ${}^{+9.08}_{-8.83}$ & ${}^{+4.73}_{-7.91}$  & 2.04& ${}^{+4.32}_{-4.32}$ & ${}^{+4.67}_{-7.78}$ \\ \hline

\end{tabular}
\end{center}
\caption{Inclusive Higgs production cross-section through gluon fusion (in $pb$) at $\sqrt{s}=8$ TeV, with pdf and scale uncertainties for the MSTW08 and ABM11 pdf sets. The pdf uncertainty for MSTW08 is calculated using the $90\%$CL grids, for reasons explained in section~\ref{sec:pdfs}, while the ABM11 uncertainty corresponds to $68\%$CL.  }
\label{xsection2}
\end{table}

\newpage

\end{document}